\documentclass[aps,prl,reprint,floatfix,showpacs,groupaddress]{revtex4-1}

\usepackage{amssymb}
\usepackage{amsmath}
\usepackage{bm}
\usepackage{graphicx}
\DeclareMathOperator{\Tr}{Tr}

\begin{document}

\title{Anisotropic disclination cores in nematic liquid crystals modelled by a self consistent molecular field theory}

\author{Cody D. Schimming}
\email[]{schim111@umn.edu}
\affiliation{School of Physics and Astronomy, University of Minnesota, Minneapolis, Minnesota 55455, USA}

\author{Jorge Vi\~nals}
\affiliation{School of Physics and Astronomy, University of Minnesota, Minneapolis, Minnesota 55455, USA}

\date{\today}

\begin{abstract}
Disclination configurations of a nematic liquid crystal are studied within a self-consistent molecular field theory. The theory is based on a tensor order parameter, and can accommodate anisotropic elastic energies without the known divergences in the Landau-de Gennes formulation. Our results agree with the asymptotic results of Dzyaloshinsky for the Frank-Oseen energy and far from the defect core, but reveal biaxial order at intermediate distances from the core, crossing over to uniaxial, but isotropic configurations as the core is approached. The elastic terms considered in our energy allow for separate control of surface tension, anchoring, and elasticity contrast, and are used to analyze recent results for lyotropic chromonic liquid crystals. The latter display unusually large defect cores (on the order of tens of microns) which can be used for a quantitative comparison with the theory. Both $\pm 1/2$ disclination configurations are well reproduced by our calculations. Elastic anisotropy is also shown to lead to qualitative changes in the disclination polarization, a quantity that is proportional to the active stress in models of active matter. 
\end{abstract}

\maketitle

The Landau-de Gennes theory, originally introduced to model the isotropic to nematic phase transition in a liquid crystal, forms the basis of many if not most theoretical and computational studies of equilibrium and nonequilibrium phenomena in systems that exhibit broken orientational symmetry \cite{deGennes75,gramsbergen86}. They range from conventional liquid crystals and liquid crystal elastomers \cite{selinger16,re:turiv20}, to active matter and self-organizing living systems \cite{marchetti13,genkin17,shankar18,binysh20}. The Landau-de Gennes theory introduces a tensor order parameter $\bm{Q}$ to describe local order, distinct from the more classical approach based on a director field $\bm{\hat{n}}$ which represents the average orientation of anisotropic molecules. In director based theories, boundary conditions and topological constraints generically lead to configurations with point or line singularities in the director field. Therefore director models must address unbounded spatial derivatives near the singularities, and, with them, diverging elastic energies \cite{frank58,selinger19}. However, for most liquid crystals the characteristic size of the physical core of the singularity is much smaller than scales of interest, so that a short length scale cut-off is often sufficient to eliminate the need to resolve the director field at too fine a scale. Since elastic constants and generalized viscosities are well characterized experimentally, the director representation with a short distance cut-off has become the theory of choice to study equilibrium and transport in macroscopic size nematic systems.

The assumption that local order is described by a director field breaks down near singularities, at two phase interfaces, and in some cases near boundaries \cite{lyuksyutov78,sen86,schopohl87,popanita97}. It is then replaced in the Landau-de Gennes theory by the symmetric, traceless tensor order parameter $\bm{Q}$ that allows for biaxial order: Molecular configurations in which the orientation cannot be described by a single direction. Order is now characterized by the eigenvalue structure of $\bm{Q}$: all eigenvalues vanish for an isotropic phase, a degenerate eigenvalue describes a uniaxial phase, and three distinct eigenvalues signal biaxial order. An important additional advantage of the Landau-de Gennes theory is that it eliminates diverging elastic energies at director field singularities as the tensor's eigenvalues change smoothly over regions in which the director changes rapidly. However, it is still necessary to resolve multiple length scales, down to the scale of director singularities. This makes it difficult to use the theory for the study of macroscopic configurations and flow. 

There are several well known difficulties associated with Landau-de Gennes theory. First, elastic constants in the director and tensor representations can be mapped into each other only away from singularities. Therefore, while the constants can be nominally related to experimentally determined elastic constants, the elastic energies near singularities of $\bm{\hat{n}}$ remain largely phenomenological. Second, to lowest (second) order in a gradient expansion of the free energy as a function of $\bm{Q}$, the elastic energy is isotropic (it does not distinguish splay and bend distortions, for example). Third order terms in gradients are needed to break this degeneracy, but it is known that the free energy at this order becomes unbounded for all values of its parameters. Therefore, the requirement of a stable free energy implies consideration of terms at least of fourth order in gradients (there are 22 possible terms allowed by symmetry \cite{longa87}), thus making the theory intractable for anisotropic systems. The lack of boundedness can be traced back to the fact that the Landau-de Gennes theory as formulated does not constrain the eigenvalues of $\bm{Q}$ to remain within their physically admissible range \cite{ball10}. The ability to extend the Landau-de Gennes theory to anisotropic systems is imperative for contemporary applications in active and living nematics \cite{marchetti13,genkin17,shankar18,aranson19,binysh20,coelho20}, in surface actuation and the study of the effects of surface curvature \cite{shin08,most15,jimenez16,baba18,nestler20}, or in transport of droplets and biological materials in nematic media \cite{peng15,peng18,hokmabad19,lee20}. 

A computational framework is introduced, based on a tensor order parameter, that accommodates elastic anisotropy and fully biaxial configurations. It avoids divergences at third order in gradients by self consistently constraining the range of variation of the eigenvalues of $\bm{Q}$. Our calculations address recent benchmark experiments on lyotropic chromonic liquid crystals that have revealed very large disclinations in the nematic phase, with cores on the order of a few microns \cite{zhou17}, and hence resolvable by optical means. The measurements show strong anisotropy in the director configuration, and also a transition from a uniaxial state far away from the core, to biaxial at intermediate distances, and back to uniaxial near the core. These experiments evidence the unique and complex interplay between elasticity, anisotropy, and topology that gives rise to the anisotropic morphology. We show that it is sufficient to consider two anisotropic gradient terms to separately control energy anisotropy and elastic constant constrast, and use the experimentally determined optical retardance to determine the values of the coupling coefficients. Good agreement with experiments is found for both +1/2 and -1/2 disclinations at all distances from the core, thus validating the tensor order parameter theory down to the core of the singularity, including anisotropic and biaxiality effects.

Our work is based on a self consistent, mean field theoretic extension of the Maier-Saupe molecular field theory \cite{katriel86,ball10,schimming20}, including anisotropic terms in the elastic free energy. We obtain equilibrium configurations $\bm{Q}(\bm{r})$ that minimize a free energy functional $F[\bm{Q}] = H[\bm{Q}] - T \Delta S$, where $H$ is the Hamiltonian of a configuration to be defined below, and $\Delta S$ is the entropy relative to the isotropic state. Unlike the Landau-de Gennes theory, we conduct a constrained minimization restricted to tensors of the form
\begin{equation} 
\label{Qdef}
\bm{Q}(\bm{r}) = \int_{S^2} \big(\bm{\hat{u}} \otimes \bm{\hat{u}} - \frac{1}{3} \bm{I}\big) p(\bm{\hat{u}},\bm{r}) \, d \bm{\hat{u}}
\end{equation}
where $\bm{\hat{u}}$ is the molecular orientation, and $\bm{I}$ is the identity tensor. Any configuration $\bm{Q}(\bm{r})$ satisfying this constraint will have eigenvalues $-1/3 \leq q \leq 2/3$. The function $p(\bm{\hat{u}},\bm{r})$ is the canonical probability distribution at fixed temperature, allowed to vary in space (via the assumption of local equilibrium). The entropy is given by $\Delta S = - n k_B \langle \ln 4 \pi p(\bm{\hat{u}},\bm{r}) \rangle$, which we compute self consistently with Eq. (\ref{Qdef}). The entropy of a given configuration $\bm{Q}$ can be obtained by maximization over microscopic configurations subject to the constraint (\ref{Qdef}). If $\bm{\Lambda}(\bm{r})$ is the associated tensor of Lagrange multipliers, the entropy is maximized by
\begin{equation} 
\label{prob}
p(\bm{\hat{u}},\bm{r}) = \frac{\exp[\bm{\hat{u}}^{T} \bm{\Lambda}(\bm{r}) \bm{\hat{u}}]}{Z[\bm{\Lambda}(\bm{r})]} 
\end{equation}
where $Z[\bm{\Lambda}(\bm{r})]$ is interpreted as a single particle partition function and $\bm{\Lambda}(\bm{r})$ an effective conjugate field. This partition function cannot be obtained analytically. The method to compute it for spatially varying configurations of $\bm{Q}(\bm{r})$, and the associated, self-consistent free energy minimization are key results of our work. Note that although the molecular units are uniaxial, when $\bm{\Lambda}$ has three distinct eigenvalues the distribution over $\bm{Q}$ is biaxial \cite{schimming20}. 

Substituting Eq. \eqref{prob} into the constraint, Eq. \eqref{Qdef}, yields the the following self-consistency relation
\begin{equation} \label{selfCon}
\bm{Q} + \frac{1}{3} \bm{I} = \frac{\partial \ln Z[\bm{\Lambda}]}{\partial \bm{\Lambda}}. 
\end{equation}
This equation serves to implicitly determine $\bm{\Lambda}$ as a function of $\bm{Q}$, which is necessary to minimize $F$. It has been shown that if the eigenvalues of $\bm{Q}$ approach the limits of the physical range, $\bm{\Lambda}$, and thus the free energy, diverges \cite{ball10}.

We choose the Hamiltonian $H[\bm{Q}] = \int_{\Omega} \{ - \alpha \Tr [\bm{Q}^2] + f_e(\bm{Q},\nabla \bm{Q})\} \, d\bm{r}$ where $\alpha$ is an interaction strength parameter. The gradient independent contribution is an extension of the original Maier-Saupe Hamiltonian \cite{maier59,selinger16} to which an elastic term $f_e$ is added to penalize spatial variation. The following functional form of the elastic free energy will be used
\begin{eqnarray} 
\label{elastic} 
f_e(\bm{Q},\nabla \bm{Q}) = L_1 \partial_k Q_{ij} \partial_k Q_{ij} & + & L_2 \partial_j Q_{ij} \partial_k Q_{ik} \nonumber\\
& + & L_3 Q_{k\ell} \partial_k Q_{ij} \partial_{\ell} Q_{ij}
\end{eqnarray}
where $\partial_k$ stands for the partial derivative $\partial / \partial x_k$,  and summation over repeated indices is assumed. With this choice of elastic energy, cubic in $\bm{Q}$, the Landau-de Gennes free energy is unbounded below \cite{longa87,ball10}. However, because of the constraint on the eigenvalues of the field $\bm{Q}$, implicit in the form of Eq. (\ref{Qdef}), the free energy resulting from the partition function $Z$ in Eq. (\ref{prob}) remains bounded within a finite range of $L_3 / L_1$ \cite{bauman16}.

We study order parameter configurations around positive and negative disclinations as in the experiments of Ref. \cite{zhou17}. In the lyotropic chromonic liquid crystals used in the experiments, it has been established that splay $K_{11}$ and bend $K_{33}$ elastic constant are different for a large range of temperatures and molecular concentrations \cite{zhou12,zhou14}. The elastic energy, Eq. (\ref{elastic}) has the fewest number of terms required to lift the degeneracy between splay and bend while allowing a degree of separate control of anchoring and elastic anisotropies. If the eigenvalues of $\bm{Q}$ are uniform, $f_e$ in Eq. (\ref{elastic}) can be mapped to the Frank-Oseen energy with $K_{33} - K_{11} = (16/3) S^3 L_3$ where $S = (3/2) q_{max}$ is the usual scalar order parameter for uniaxial nematic liquid crystals, and $q_{max}$ denotes the largest eigenvalue of $\bm{Q}$. In the experiments performed in thin films, the director is fixed in the $xy$ plane, and the twist term in the Frank-Oseen energy is always zero.

\begin{figure}
	\includegraphics[width = \columnwidth]{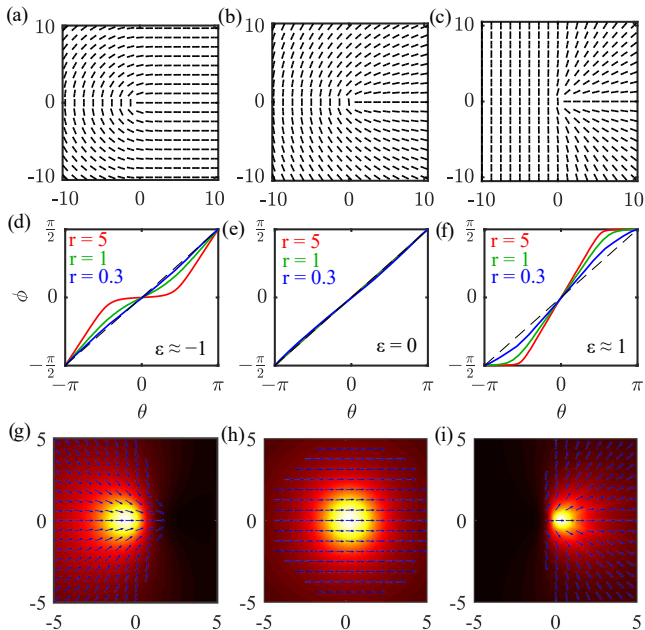}
	\caption{(a-c) Director configuration, (d-f) director angle $\phi$ versus polar angle $\theta$ for three radial distances from the core, and (g-i) disclination polarization $\nabla \cdot \bm{Q}$. Left: $\varepsilon \approx -1$, center: $\varepsilon = 0$, and right: $\varepsilon \approx 1$. For $\varepsilon \neq 0$ the polarization is nonuniform and exhibits contractile or extensile vector flows.}
	\label{fig:dzyal}
\end{figure}

Dimensionless variables are introduced as follows $\tilde{f} = f/(n k_B T)$,  $\tilde{x} = x / \xi$, $\xi = \sqrt{L_1 / (n k_B T)}$, $\tilde{L_2} = L_2 / L_1$, $\tilde{L_3} = L_3 / L_1$ where $f$ is the free energy density. We subsequently drop the tilde for brevity. We find the minimum of the free energy by solving the Euler-Lagrange equations, $\delta F / \delta \bm{Q} = 0$, given uniaxial outer boundary conditions for $\bm{Q}$ with the specified degree of the singularity, and by solving numerically the self consistency equations involving $\bm{Q}(\bm{r})$ and $\bm{\Lambda}(\bm{r})$, Eq. \eqref{selfCon} \cite{schimming20}. For all computations $\alpha / (n k_B T) = 4$ so that the system is below the supercooling limit and the nematic phase is the stable phase with an equilibrium $S = 0.6751$. To analyze the results, we parametrize solutions by $\bm{Q} = S(\bm{\hat{n}} \otimes \bm{\hat{n}} - (1/3) \bm{I}) + P(\bm{\hat{m}} \otimes \bm{\hat{m}} - \bm{\hat{\ell}} \otimes \bm{\hat{\ell}})$ where $P$ is the degree of biaxiality of the distribution, $\bm{\hat{n}}$ is the director, and $\{\bm{\hat{n}},\bm{\hat{m}},\bm{\hat{\ell}}\}$ form an orthonormal frame. $P$ can be calculated from the difference of the two smallest eigenvalues of $\bm{Q}$. We include biaxiality in the parameterization as there is no uniaxial restriction on $\bm{Q}$ except at the boundary.

We first compare our results with results given by the Frank-Oseen model with $K_{11} \neq K_{33}$ away from the core of a disclination. This problem was originally solved by Dzyaloshinsky, where solutions are parameterized by $\varepsilon = (K_{33} - K_{11}) / (K_{33} + K_{11})$ \cite{dzyaloshinsky70,hudson89,zhou17}. With this definition, in the limit $\varepsilon = -1$ the bend elastic constant goes to zero, while for $\varepsilon = 1$ the splay constant is zero. The qualitative effect of $\varepsilon$ on the spatial order parameter profile is easiest to visualize for $+1/2$ disclinations since in the case of zero bend constant, the director exhibits distortions with no splay and vice versa for the case of zero splay constant. For $-1/2$ disclinations the director cannot relax to remove all of one type of distortion. In Figs. \ref{fig:dzyal}a-c we show the director configurations given by the molecular field theory studied here for the cases of $\varepsilon \approx -1$, $\varepsilon = 0$, and $\varepsilon \approx 1$ for a $+1/2$ disclination. In Figs. \ref{fig:dzyal}d-f we show the angle of the director with respect to the $x$ axis $\phi$ versus the polar angle $\theta$ for various distances away from the core. Away from the core, the director conforms to the Dzyaloshinsky solution, as seen by the deviation from $\phi = (1/2)\theta$. Near the core, the director profile becomes isotropic (i.e. $\phi(r \to 0) \sim (1/2) \theta$) which can be understood by the fact that $S$ becomes small in this region, and the anisotropic elastic constant difference, $K_{33} - K_{11} \propto S^3$, is negligible. Our solutions of the molecular field theory smoothly interpolate between these two asymptotic behaviors. 

Figs. \ref{fig:dzyal}g-i show the computed disclination polarization $\nabla \cdot \bm{Q}$ for the $+1/2$ disclination. This quantity is important in the field of active matter as it is  proportional to the active force \cite{ramaswamy10,doo18,aranson19}. Such an active force is argued to lead to flows and motion of $+1/2$ disclinations, while $-1/2$ disclinations remain stationary. Active matter models to date have neglected elastic anisotropy, and focused on the  \lq\lq one constant approximation'' \cite{shankar18,binysh20}. Our solutions to the molecular field theory show that although the director profile becomes isotropic at the center of the core, its anistropy results in a nonuniform $\nabla \cdot \bm{Q}$ across the disclination. These nonuniformities would result in qualitatively different, spatially nonuniform active forces acting on $+1/2$ disclinations; in particular, contractile forces for $\varepsilon \approx -1$ and extensile forces for $\varepsilon \approx 1$ (or vice versa if the proportionality between the active force and $\nabla \cdot \bm{Q}$ is negative).

\begin{table*}
\caption{\label{eTerms} Expansion of terms in the elastic energy, Eq. \eqref{elastic}.}
\begin{ruledtabular}
\begin{tabular}{c c c c}
Term & $f_{\sigma}$ & $f_{w}$ & $f_{FO}$ \\ \hline
$\partial_k Q_{ij} \partial_k Q_{ij}$ & $\frac{2}{3}|\nabla S|^2$ & $0$ & $2 S^2 \big( (\nabla \cdot \bm{\hat{n}})^2 + |\bm{\hat{n}} \times (\nabla \times \bm{\hat{n}})|^2\big)$ \\
$\partial_j Q_{ij} \partial_k Q_{ik} $ & $\frac{1}{9}|\nabla S|^2$ & $\frac{1}{3} (\bm{\hat{n}} \cdot \nabla S)^2 + \frac{2}{3} S(\bm{\hat{n}} \cdot \nabla S)(\nabla \cdot \bm{\hat{n}}) + \frac{2}{3} S \{ \nabla S \cdot \big(\bm{\hat{n}} \times (\nabla \times \bm{\hat{n}})\big) \} $ & $ S^2\big((\nabla \cdot \bm{\hat{n}})^2 + |\bm{\hat{n}} \times (\nabla \times \bm{\hat{n}})|^2\big)$ \\
$Q_{k\ell} \partial_k Q_{ij} \partial_{\ell} Q_{ij}$ & $-\frac{2}{9} S |\nabla S|^2$ & $\frac{2}{3} S (\bm{\hat{n}} \cdot \nabla S)^2$ & $2 S^3\big( |\bm{\hat{n}} \times (\nabla \times \bm{\hat{n}})|^2 - \frac{1}{3}(\nabla \cdot \bm{\hat{n}})^2\big)$ 
\end{tabular}
\end{ruledtabular}
\end{table*}

The consequences of the gradient terms in the elastic energy can be qualitatively understood by considering a uniaxial solution of the form $\bm{Q}(\bm{r}) = S(\bm{r}) (\bm{\hat{n}}(\bm{r}) \otimes \bm{\hat{n}}(\bm{r}) - (1/3) \bm{I})$. Substituting this into Eq. \eqref{elastic} explicitly separates contributions to gradients in the degree of ordering $S$ and the director $\bm{\hat{n}}$. We note three types of contribution to the energy: a surface tension term $f_{\sigma}(S,\bm{\hat{n}})$, an anchoring energy term $f_{w}(S,\bm{\hat{n}})$ (i.e. an energy that depends on the relative orientation of $\bm{\hat{n}}$ to $\nabla S$), and $f_{FO}(S,\bm{\hat{n}})$, an elastic contribution of the Frank-Oseen type. Note that although we do not have a two phase interface or solid surface, the underlying order changes rapidly and it is useful to name the respective terms as surface tension and surface anchoring as they would appear on interfaces. These three contributions to the elastic free energy are summarized in Table \ref{eTerms}. Larger surface tension contributions lead to larger disclinations, but do not contribute to the anisotropy in the morphology of the core. Anchoring energies, on the other hand, directly result in anisotropic shapes through its dependence on the angle between $\bm{\hat{n}}$ and $\nabla S$, while contributions in $f_{FO}$ only result in anisotropy if the bend and splay degeneracy is broken.

We finally compare our solutions corresponding to the disclination configurations of Ref. \cite{zhou17}. As appropriate for the experiments, we present results for the optical retardance, $\Gamma = \gamma(S - P)$. The results are reported as a function of angular Fourier modes: $\Gamma(r,\theta) = \Gamma_0(r) + \Gamma_1(r) \cos\theta + \Gamma_3(r) \cos 3\theta$. The higher order Fourier modes are thus measures of anisotropy of the ordering fields of the solution. Elastic energies with no anchoring contribution, $f_w$ = 0, and degenerate Frank elastic constants lead to an isotropic core with $\Gamma(r,\theta) = \Gamma_0(r)$. Thus, both anisotropic terms in Eq. \eqref{elastic} will yield nonzero higher order Fourier modes in $\Gamma$. However, as seen in Table \ref{eTerms}, the $L_2$ term also has a positive surface tension contribution, which yields larger disclinations as it is increased and the anisotropy becomes less pronounced. On the other hand, the $L_3$ term does not contribute to the size of the disclination due to the negative surface tension term and the anisotropy of disclinations is sensitive to this term throughout the valid range of $L_3$. However, if $L_3$ is chosen so that $|\varepsilon| > 1$ the energy becomes unstable. Because larger values of $L_2$ decrease $\varepsilon$ (because they increase the overall magnitude of $K_{11}$ and $K_{33}$), $L_2$ and $L_3$ can be adjusted to find the desired anisotropy (e.g. peak height of $\Gamma_{i\neq0}$) while also conforming to the physically relevant value of $\varepsilon$.  

\begin{figure}
	\includegraphics[width = \columnwidth]{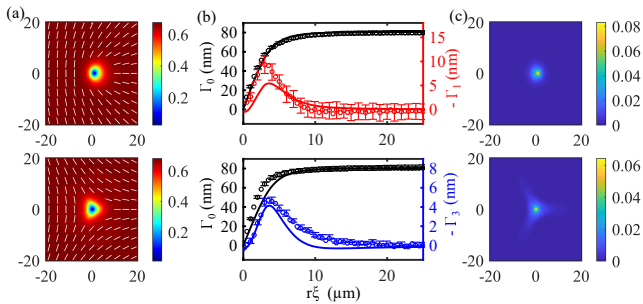}
	\caption{(a,b) Optical retardance $\Gamma$ and (c) biaxiality $P$ for a $+1/2$ disclination (top) and $-1/2$ (bottom). Solid lines in (b) are solutions to the molecular field theory, dots are the experimental data points in Ref. \cite{zhou17}.}
	\label{fig:discs}
\end{figure}

In Fig. \ref{fig:discs} we show our computational results for the spatial profile of the optical retardance and the biaxiality $P$ for $\pm 1/2$ disclinations with $L_2 = 7$ and $L_3 = 5$. In order to compare with experiments, we plot the angular Fourier modes along with the experimental data from Ref. \cite{zhou17}, Fig. \ref{fig:discs}b. As in the experiments, we have determined $\gamma$ and $\xi$ by fitting the computed $\Gamma_0$ to the experiment for the $+1/2$ disclination. We find $\gamma = 119.75 nm$ and $ \xi = 1.2 \mu m$. The values of $L_{2}$ and $L_{3}$ chosen are not independent; rather they satisfy $\varepsilon = 0.4$, as appropriate for the chromonic liquid crystal used \cite{zhou14,zhou17}. We find that the anisotropy, quantified by the angular Fourier modes is peaked roughly at the ``edge'' of the disclination core, while it goes to zero far from the core and at the core center. Importantly, even though we use an elastic energy with the minimal number of free parameters needed to break the Frank degeneracy, and have only one free parameter to determine, the model agrees with experiments for both $\pm 1/2$ disclinations.

We also find anisotropy in the biaxiality, as shown in Fig. \ref{fig:discs}c. While subtle for the $+1/2$ disclination, it is prominent for the $-1/2$ disclination, and extends away from the core into the regions where the director is radially oriented relative to the disclination center. This can be understood by the fact that the splay constant is reduced, and, thus, it is more favorable for the distribution to become biaxial in the splay dominated regions. However, close to the core center the biaxiality becomes isotropic, similarly to $\Gamma$. This is likely due to the reduction of the eigenvalues of $\bm{Q}$ near the core as discussed previously. We note that the appearance of biaxiality has been studied in isotropic disclination cores by using a Landau-de Gennes free energy \cite{lyuksyutov78,meiboom82,schopohl87}. In our calculations, the biaxiality is a result of the single particle partition function instead, Eq. \eqref{prob}, since the Hamiltonian itself only favors uniaxial configurations.

In summary, we have developed a molecular field theory of nematic order in liquid crystals that includes anisotropy and elasticity contrast while yielding stable results. Further, anisotropic disclination cores - in both eigenvalues and eigenvectors of $\bm{Q}$ - result from such an elastic free energy, with the results for the order parameter configuration quantitatively agreeing with experiments, even for the minimal number of parameters required. We have also introduced a simple conceptual framework to understand the contribution of these elastic free energies to the structure of defects, and shown that terms contributing to the classic Frank-Oseen elastic energy lead to anisotropic morphologies when $K_{11} \neq K_{33}$. Our results are also relevant for models of active matter as anisotropic energies would result in disclination configurations with qualitatively different spatial distributions of active forces, as shown in Fig. \ref{fig:dzyal}(g)-(i). In addition, the interplay between anchoring and elasticity anisotropy on the one hand, and surface tension on the other, will be important in the study of two phase domain coexistence (tactoids), their interaction, and structure coarsening.


We thank Sergij Shiyanovskii for stimulating discussions, and him as well as the other authors of Ref. \cite{zhou17} for sharing the raw data of the experiments with us. This research has been supported by the National Science Foundation under Grant No. DMR-1838977, and by the Minnesota Supercomputing Institute.

%

\end{document}